\documentclass[suppldata]{interact}

\usepackage{epstopdf}
\usepackage[caption=false]{subfig}

\usepackage[natbibapa,nodoi]{apacite}
\usepackage[utf8]{inputenc}
\usepackage{amsmath}
\usepackage{hyperref}
\usepackage{url}
\usepackage{graphicx}
\usepackage{booktabs}
\usepackage{lipsum}

\theoremstyle{plain}

\theoremstyle{definition}

\theoremstyle{remark}

\begin{document}

\title{On computational approaches to Pop music culture}

\author{
\name{A. Flexer\textsuperscript{}\thanks{CONTACT A. Flexer. Email: arthur.flexer@jku.at} }
\affil{\textsuperscript{}Institute for Computational Perception, Johannes Kepler University Linz, Austria}
}

\maketitle

\begin{abstract}
This overview article presents arguments why the computational study of Pop music culture needs to be conducted in a multi-modal way beyond mere audio analysis, gives a survey of already published quantitative work on analyzing Pop music at scale, and discusses challenges and promising research avenues for future work. 

We argue that Pop music culture is a rich tapestry of audio, visual, textual and cultural connotations and relations which needs to be studied in an integrative way as a multi-modal socio-cultural phenomenon. What is needed is an approach which is reminiscent of "distant reading", i.e.\ algorithmic analysis of thousands of books as a research tool in digital humanities. In addition to listening to audio, algorithms need to view album artwork and music videos, to read meta-information, lyrics, music magazines and books.

Our review of already available work on distant reading/listening/viewing and multi-modal combinations thereof reveals two major open issues: a scarcity of truly multi-modal approaches and questionable external validity rooted in sampling practices when building music corpora. In trying to overcome these shortcomings we sketch three exemplary avenues for future research on Pop music culture: charting the topic universe of music lyrics, providing an iconography of album cover art, tracking retro cycles in music’s timeline.

\end{abstract}


\section{Introduction}
\label{introduction}

\begin{figure}[!htbp]
\centering
\subfloat{%
\resizebox*{5cm}{!}{\includegraphics{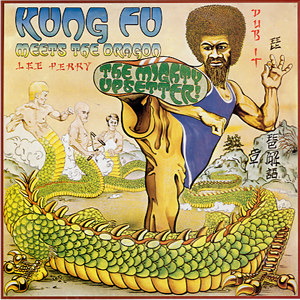}}}\hspace{50pt}
\subfloat{%
\resizebox*{5cm}{!}{\includegraphics{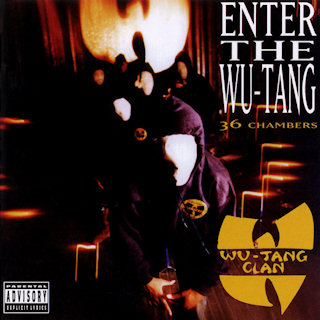}}}
\caption{Two classic Pop music albums by Lee Perry (left) and the Wu-Tang Clan (right). Images taken from Wikipedia as fair use.} \label{sample-figure}
\label{fig:perry_tang}
\end{figure}

It is the goal of this survey article to present an overview of the application of music information retrieval (MIR) methods to the study of Pop music culture in all its multi-modal facets at large scale. MIR is the interdisciplinary science of retrieving information from music, using a multitude of methods from signal processing, statistics, machine learning, artificial intelligence, etc (see \cite{peeters2025} for a recent overview). For this article we adopt the constructive definition of "Pop" as any type of music that a person has been exposed to by the mass media \citep{Boyle:etal:1981}. Let us give a motivating example making it clear that "pop music is only partly music"  \citep[p.~1]{Diederichsen:2023} and should be understood as  a multi-modal socio-cultural phenomenon. Take the two classic Pop music albums shown in Figure \ref{fig:perry_tang}:  "Kung Fu Meets the Dragon" (1975) by Lee Perry and "Enter the Wu-Tang (36 Chambers)" (1993) by the Wu-Tang Clan. Lee Perry is not only seen as the founder of Reggae and Dub music, but as a pioneer of studio recording technology essentially inventing remixing and sampling, thereby influencing countless musicians ranging from Rock to Techno to Hip-Hop. The album itself is inspired by the wave of martial arts movies that were popular at the time of recording, both concerning the cover artwork and the use of Chinese sound effects citing Lalo Schifrin's 1973 movie soundtrack "Enter The Dragon". The Hip-Hop album by the Wu-Tang Clan refers to this kind of Buddhist philosophy taken from Kung Fu movies, it uses many of the production tricks pioneered by Lee Perry and even the album artwork connects to its predecessor with its use of faux Chinese typography. 

The above example is an illustration that "Pop music is a constant, virtual amalgamation of diverse media and sign systems from the real world" \citep[p.~38]{Diederichsen:2023}. In order to do justice to this rich tapestry of audio, visual, textual and cultural connotations, MIR should study Pop music as a multi-modal socio-cultural phenomenon. In order to do this at large scale, MIR needs an approach which is reminiscent of "distant reading", i.e.\ algorithmic analysis of thousands of books as a research tool in digital humanities \citep{Moretti:2000,Moretti:2005}. In addition to "distant listening" \citep{have2021close} to audio, MIR algorithms need to "distant view" \citep{arnold2019distant} album artwork and music videos, as well as to "distant read" meta-information, lyrics, music magazines and books.  

The remainder of this article is divided into two parts: 

\begin{itemize}
    \item providing a substantial review of already existing work on computational analysis of Pop music at scale in section \ref{stateoftheart}
    \item identifying open issues and drafting exemplary research questions concerning Pop music culture that could be approached in a data driven manner in section \ref{agenda}
\end{itemize}

This article therefore discusses to what extent MIR is able to provide empirical results and evidence concerning Pop music culture, thereby adding to existing theories from humanities concerned with Pop music. It will therefore help to  consolidate a computational and digital humanities approach to Pop music culture.

\section{State of the Art}
\label{stateoftheart}

In this section we will first give a brief overview of Pop music theory and distant reading in the humanities, since these are two fields outside of MIR which offer important insights for our review. Next we discuss the state of the art concerning distant reading/listening/viewing in MIR, plus multi-modal approaches combining more than one modality. Our survey will put an emphasis on large scale MIR results concerning Pop music culture where available, i.e.\ corpora of thousands rather than hundreds of songs. Otherwise we report about results and methods most closely related to this focus of our article.

Other interesting overview articles on multi-modal MIR not necessarily centering on Pop music culture include an advocation for multi-modal and user-centered strategies \citep{liem2011need}, a report about a week long seminar on multi-modal music processing \citep{muller2011multimodal}, and appraisals and guidelines concerning multi-modal music datasets \citep{christodoulou2024multimodal,gotham2025towards}. These references provide excellent additional reading beyond the computational musicology focus of this survey.

\subsection{Pop music theory}

"Pop" as a genre originated in its modern form during the 1950s in the USA and UK, at the beginning being mostly synonymous with Rock'n'Roll but today encompassing popular music of a wide range of styles. It has often been defined negatively, e.g.\ as being different from jazz and folk music \citep{hatch1987blues} or as non-classical music. Another more constructive definition is any type of music that a person has been exposed to by the mass media \citep{Boyle:etal:1981}, with its exact definition itself being subject to constant change and evolution.

Pop's main musical form is the song, often of rather short duration, having been described as containing noticeable rhythmic elements, a mainstream style and a rather simple structure of verse and chorus \citep{everett2000expression}. Apart from the musical form, innovations in media technology (inexpensive durable 45rpm vinyl record, portable transistor radio, television, music video) have shaped Pop music into a multi-modal socio-cultural phenomenon.

The theoretical debate of Pop music can be traced back to analysis of general aspects of mass culture in the 1940s \citep{adorno1941popular}, followed by an orientation towards more popular forms of mass culture in the 1950s \citep{Rosenberg:White:1957}, leading to founding of the "Popular Culture Association" in the USA in 1969 and their "Journal of Popular Culture"\footnote{\url{http://www.journalofpopularculture.com/}}, succeeded by the journal "Popular Music"\footnote{\url{https://www.cambridge.org/core/journals/popular-music}} in 1981. Also still relevant today are "culture analytic" approaches in the tradition of "cultural studies", focusing on roles of individuals in forming subcultures \citep{Bennett:2000} or on cultural production of subjectivity \citep{Schwarz:1997} influenced by Jacques Lacan's semiotic theories. More recent strands of "new musicology" \citep{Kerman:1985} emphasize the intimate relationship between music and society and how music participates in social formation of individuals, thereby employing methods from anthropology, sociology, cultural studies, gender studies and feminism. 

\subsection{Distant reading in digital humanities}

The accessibility of vast amounts of text in digital form has enabled humanities to add "distant reading" of thousands of books via computational analysis as a new research tool to its repertoire of methods. The term distant reading is attributed to an article by \cite{Moretti:2000} and has been further popularized by his influential book \citep{Moretti:2005}. Distant reading is usually applied to large collections of text often of a magnitude which cannot be handled by individual scholars in what is known as traditional "close reading", i.e.\ very careful and detailed expert reading of only comparably few texts. Distant reading relies heavily on empirical, quantitative and algorithmic methods, thereby being able to handle large amounts of texts, but at a coarser level of detail than close reading approaches.  To give one example, the digitization effort by the "Google Books Team" enabled computational analysis of more than 5 million books \citep{michel2011quantitative}, comprising about $4\%$ of all books ever printed at that time. This made it possible to quantify phenomena like e.g.\ peak usage of the word "slavery" during the US civil war (1861–1865) and the civil rights movement (1955–1968) via observation of word frequencies over the years. Similar to tracing the rise and oblivion of topics one can apply the same principle to chart the influence of artists, politicians, scientists, etc over time. Such an application of high-throughput data collection and analysis to the study of human culture has been termed "Culturomics" \citep{michel2011quantitative}. Similar "distant" approaches have been explored for listening to audio \citep{have2021close} or viewing of images \citep{arnold2019distant}.

The methods and the overall intention of distant reading are sometimes discussed controversially in the humanities, with some fearing that it might replace close reading altogether \citep{ascari2014dangers}, or others pointing out that the capabilities of the algorithmic tools might dictate what hypotheses can be formulated and proven \citep{fish2012mind}, or that researchers might be tempted to formulate conjectures after looking at the data, thereby preventing usage of that data to test validity of these post hoc theories \citep{huron2013virtuous}. We endorse the constructive point of view that close and distant reading should not be seen as competitors but as complimentary approaches excelling at different scales of analysis \citep{coles2013solitary}. Therefore distant methods should always be complemented with traditional close studying of sources. 

\subsection{Distant reading in MIR}

When developing and validating hypotheses in musicology, relevant information very often is obtained from written documents. This information from collections, anthologies, compilations, biographies, reviews, journals, etc is today often available in digitized formats, enabling usage of methods from natural language processing (NLP) for music knowledge discovery \citep{Oramas:etal:2018}. For the field of MIR, in 2020, 2021 and 2024 there even existed a dedicated workshop on "NLP for Music and Audio"\footnote{NLP4MUSA: \url{https://sites.google.com/view/nlp4musa} and \url{https://sites.google.com/view/nlp4musa-2024/}}, co-located with MIR's major conference ISMIR\footnote{\url{https://ismir.net/conferences/}}.

At the beginning of many NLP approaches is the derivation of structured meta-data (e.g.\ knowledge graphs) from unstructured or semi-structured sources (e.g.\ texts from the internet). One example is work on
automatic band member detection and automatic recognition of all their released records from texts obtained via web crawls or directly from sources like Wikipedia or allmusic\footnote{\url{https://www.allmusic.com/}} \citep{Knees:Schedl:2011}. This approach is strongly related to "named entity detection", i.e.\ the recognition of persons’ names in documents in NLP. The overall result in this study was that manually generated rules still outperformed supervised learning approaches in retrieving band members and released records.

In a survey article \citep{Oramas:etal:2018} on using NLP for music knowledge discovery in MIR, the authors not only describe typical NLP processing pipelines but also report results on flamenco, Renaissance and Pop music. 
In a sentiment analysis of 263,525 Pop music reviews, results show more positive sentiments in reviews around 2008, especially for genres from more diverse communities like Jazz and Latin music, but not for Country music. This is attributed to economical and geopolitical circumstances like the 2008 election of US president Obama. Other more detailed analysis e.g.\ shows more positive sentiment of reviews for Reggae music between 1975 and 1985, which concurs with the so-called "golden age" of Reggae.

Related work on non-Pop genres includes discovery of social and professional
networks from Wikipedia articles on Renaissance musicians \citep{Fujinaga:Weiss:2016}, extraction of semantic information from an online
discussion forum on Carnatic music \citep{Sordo:etal:2012}, or very detailed cross-linking of references  to  musical  passages  in musicological texts \citep{Sutcliffe:etal:2015}.

The lyrics of music with singing voices are a special kind of important textual information, since they are able to convey emotional expressions, topics, and stories which can greatly influence a listener's impression of songs. In a comprehensive survey  \citep{Watanabe:Goto:2020} the field of lyrics-related studies has been termed "lyrics information processing" (LIP), sharing core technologies with both NLP and MIR. Important aspects of LIP include lyrics structure analysis
(e.g.\ rhyme scheme identification \citep{Addanki:Wu:2013} or verse-bridge-chorus labeling \citep{Mahedero:etal:2005}) and lyrics semantic analysis. The latter approach includes estimating the mood or emotion of lyrics \citep{Hu:Downie:2010,Delbouys:etal:2018,Mishra:etal:2021}, or modeling different topics in lyrics \citep{Kleedorfer:etal:2008,Sasaki:etal:2014,Sterckx:etal:2014}, e.g.\ concerning how typical they are for certain genres
\citep{aljanaki2024genre}. Whereas most of this research focuses on practical applications, e.g.\ using lyrics for new retrieval interfaces \citep{Sasaki:etal:2014}, some studies have used large-scale lyrics analysis to explore the cultural evolution of popular music. An analysis \citep{parada2024song} of more than 350,000 English songs lyrics from 1970 to 2020 showed a decrease in lexical and structural complexity as well as a trend towards more negative emotions described by the lyrics. This change towards negative sentiments had already been reported earlier \citep{napier2018quantitative} for a dataset of 6,150 "Billboard Hot 100" songs from the years 1951 to 2016 and a larger dataset of more than 150,000 songs from 1965 to 2010 \citep{brand2019cultural}. A topic analysis of $1364$ lyrics of Hip-Hop songs during the so-called ``rap wars" ($1986-1998$ in the United States) extracted two main topics, namely ``Street Life and Rhythm" and ``Vulgarity and Violence" \citep{boros2025latent}. While the former topic is more aligned with East coast Hip-Hop, the latter is more frequently used by West coast Hip-Hop artists.

Another source of textual information are microblogs and social media where users post about their music listening behaviour. "The Million Musical Tweets Dataset" \citep{Hauger:etal:2013} is an aggregation of listening histories inferred from Twitter\footnote{\url{https://twitter.com}} messages spanning 500 days, annotated with temporal and spatial information and containing pointers to artist and track details. Initial results e.g.\ showed temporal listening patterns or differences in genre preference for various countries. A related approach is to assemble music listening events directly from streaming services like "Last.fm", creating a massive data set of more than 27 billion time-stamped logs \citep{gabriel_vigliensoni}.

As a very recent development, large language models (LLMs) can be seen as a new tool for distant reading since they provide a conversational interface to enormous quantities of text. First studies on using LLMs in musicology have however mainly pointed out problems due to LLM's opaque black box nature \citep{arthur_flexer_2024_14877431}, proneness to hallucinate wrong facts \citep{ramoneda-etal-2024-role} and difficulties in logical reasoning \citep{Zhou2024CanL}. A study on using LLMs for standard MIR tasks like beat tracking, chord extraction, and key estimation on symbolic notations yielded more hopeful results \citep{fang2025}.

\subsection{Distant listening in MIR}

The concept most related to a distant listening approach in MIR are corpus studies, i.e.\ analysis of large music corpora made possible through recent advances in MIR (see \cite{Panteli:etal:2018} for an overview). These corpus studies have been reported for a variety of musics including Jazz  \citep{Shanahan:etal:2012}, Western classical music \citep{Zivic:etal:2013} and world music \citep{Panteli:etal:2018}. We now review a number of representative and exemplary studies focused on Pop music.  

In a large scale analysis of 464,411 Western Pop music songs from the years 1955 to 2010, the historical evolution of their structural regularities concerning e.g.\  pitch, timbre, and loudness have been explored \citep{Serra:etal:2012}. Using methods from statistical physics and complex networks, the authors were able to show that the variety of pitch progressions, corresponding to  the harmonic content of songs (including chords and melody), decreased over the years. Also the timbral palette became more homogeneous with already frequent timbres becoming even more frequent. The average loudness levels also increased with time, which has been attributed to the so-called "loudness war" \citep{vickers2010loudness}.

The results concerning pitch variety have been corroborated in a large scale study on vocal trends in 145,912 vocal tracks of popular songs from 1955 to 2010 \citep{georgieva2024changing}. Using source separation to extract the vocal stem and fundamental frequency (f0), a number of pitch characteristics were computed. Whereas mean pitch increased by approximately one percent per year, there also is a significant negative correlation between total variation as well as pitch class entropy and year, indicating that vocals are getting less complex over time.

In a related study on approximately 17,000 recordings from the "US Billboard Hot 100" charts between 1960 and 2010 \citep{Mauch:etal:2015} the change of the harmonic and timbral content of the music was charted using methods from evolutionary biology. Whereas other areas of the humanities like linguistics \citep{Steele:etal:2010} or archaeology \citep{Mesoudi:2011} often apply such biological tools, this approach has only recently gained traction in musicology \citep{Savage:2019}, going beyond previous somewhat controversial projects \citep{Lomax:Berkowitz:1972}.  
Aggregating both harmonic and timbral information to so-called topics the authors showed that e.g.\ the use of dominant-seventh chords, typical for Jazz and Blues music, constantly decreases over the entire observation period. On the other hand they observe a sharp increase of a timbral topic described as "energetic, speech, bright" starting in the early 1990s due to the rise of Hip-Hop music. Contrary to previous results \citep{Serra:etal:2012}, their analysis in terms of musical topics suggests that variety of music did not decrease over the years, but that Pop music evolves both continuously and in disruptions due to three stylistic "revolutions" around 1964, 1983 and 1991.

Another study \citep{Schellenberg:2012} analysed musical cues to happiness (fast tempo, major mode) and sadness (slow tempo, minor mode) in a smaller sample of about 1,000 Top 40 recordings spanning five decades. The authors report an increase in the use of minor mode and a decrease in average tempo, concluding that popular music became more sad-sounding over time. 

While music features like onset times, length of notes and melody can be extracted from audio, they are also directly represented in musical scores or other symbolic representations. However, the lack of available high-quality symbolic Pop music data has so far limited studies to only a few hundreds songs. The largest published symbolic Pop music study analyzed 1131 MIDI files containing the main melodies from the top 5 Billboard songs from 1950 to 2023 \citep{hamilton2024trajectories}. The authors describe three melodic revolutions corresponding to decreases in melodic complexity in line with audio-based results already described above \citep{georgieva2024changing}. Another notable examples identified musical phrases based on repetition in MIDI representations of 909 Chinese Pop songs \citep{DaiZD20}, and compared chord progressions in expert harmonic annotations of classical music and 921 Pop songs \citep{Sears04052021}. Another study circumvented the lack of symbolic annotations via automated melody transcription to produce MIDI representations for a corpus of 1571 Billboard chart songs, thereby uncovering a slight trend towards increasing repetitions in melodies  \citep{clark2023melody}. This new dataset is an extension of the so-called CoCoPoPs corpus \citep{arthur2023coordinated}, which itself builds on previously published harmony and melody annotations \citep{burgoyne2011expert,Temperley01092013}.

Quantitative approaches to cultural change in music have been criticized for containing specific sampling biases \citep{Fink:2013,Savage:2019}. For example two large scale studies described above \citep{Serra:etal:2012,georgieva2024changing} rely on data from the "Million Song Dataset" (MSD) \citep{Bertin:etal:2011}, which is proprietary and actually does not contain audio data but only features derived from it. Since these MSD features cover only limited aspects of music signals, e.g.\ omitting rhythm information altogether, any analysis based on them necessarily has only limited explanatory power. Other critique concerns specific methodological problems, e.g.\ casting doubt \citep{Underwood:etal:2016} on how periods of stylistic revolutions have been discovered in previous studies \citep{Mauch:etal:2015}.

\subsection{Distant viewing in MIR}

An important source of information to place an artist into a musical context is their image, be it in the form of album covers, promotional photos or music videos. After all many genres of Pop music are also defined through a certain visual aesthetic, e.g.\ heavy metal music with black clothes and long hair of the artists and very specific "dark" topics and images of the corresponding album covers. 

It has been shown \citep{Libeks:Turnbull:2011} that a simple computer vision system is able to predict music genre tags from cover artwork or promotional photos to a certain extent. Other work has documented genre differences in cover art work in terms of simple image features \citep{som,album_relationship}. An analysis of US chart music from 1945 to 2003 applied zero-shot object detection to 3130 album covers giving an overview of what different objects are depicted on album covers belonging to different music genres and types of artists \citep{lopez:flexer:2025}.

Professional music video production, which started around the early 1980s, is now an integral part of an artist's image \citep{Frith:etal:2005}. 
There exists a forum and conference series on video retrieval evaluation (TRECVID\footnote{\url{https://trecvid.nist.gov/}}), but music video analysis is still a niche topic in MIR. As one of the few exceptions, "The Music Video Dataset" comprised of more than 2000 videos from a range of genres has been published \citep{Schindler:Rauber:2016}. For copyright reasons, only Youtube-links and a number of standard visual and acoustic features are being
provided. The authors are able to estimate so-called visual object vocabularies  and their frequency distribution over video frames. This is used to capture the semantic and genre stereotypical information of music videos, e.g.\ cowboy hats, pickup trucks and acoustic guitars for Country music versus brassiere, lipstick and bikini for Dance music. 
Another study \citep{Pretet:etal:2021} explored the relationship between music and video temporal organization, showing that editors favor the co-occurrence of music and video events using strategies such as anticipation. The amount of co-occurrence depended on the music genres, e.g.\ for highly rhythmic genres like R\&B or Reggaeton many video shot transitions occurred at the downbeat positions.

\subsection{Multi-modal approaches}
\label{sec:multi_modal}

Combinations of different modalities to gain musicological insights are quite rare, with most approaches aiming for practical applications like genre classification.

Audio and text from song lyrics provide complementary information as is evident from successful combination of both modalities for automatic classification of genre, mood and emotion (see e.g.\ \citep{neumayer2007integration,KimSMMRSST10,Delbouys:etal:2018}). In terms of musicological insights, systematic connections between melodic patterns and lyrics have been explored for a rather small data set of less than 700 songs, showing that salient notes are aligned with salient parts of lyrics \citep{nichols2009relationships}. A large scale study on 119,664 lyric/audio pairs \citep{McVicarFB11} used Canonical Component Analysis to uncover correlations between linear combinations of lyrical and audio features corresponding to known aspects of mood and valence and arousal. In a study \citep{czedik2024charting} on 124,288 lyrics from the metal genre it was shown that perceived musical hardness measured with an audio feature model correlates with particularly brutal lyrics also dealing with topics like dystopia, archaisms, occultism, religion, satanism, battle and madness. A recent study \citep{Oramas2025} of more than 21,000 songs from the Billboard Hot 100 charts from 1958 until 2022 has been annotated by experts with 58  attributes mostly describing musical aspects (rhythm, compositional focus, harmony, instrumentation, sonority, vocals) but also lyrics related facets (angry, sad, happy, humorous, explicit, etc). The main results show three major stylistic revolutions in 1964, 1983, and 2016, plus two minor ones in 1991 and 2007, in accordance with previous results \citep{Mauch:etal:2015}.

A recent development are "Audio LLMs" extending pre-trained large language models (LLMs) with audio information by including tokens from audio encoders. These models have been used for automatic music tagging and cross-modal retrieval \citep{HuangJLGLE22}, automatic lyrics interpretation \citep{ZhangJXD22}, music understanding and reasoning \citep{llark2024}, captioning and question answering \citep{liu2024music,deng2024musilingo}. It will be interesting to see future application of this methodology to musicological questions and whether Audio LLMs are also prone to hallucinate facts, as first results seem to suggest \citep{WeckMBQFB24}.

Going beyond combination of only two modalities, a multi-modal embedding model trained on audio tracks, text reviews and cover art images has been used for genre classification \citep{Oramas:etal:2018b}. Providing some musicological insight, the authors used so-called 'heatmaps' \cite{Zhou:etal:2016} to visualize areas of album covers which are important for a classification decision, observing that the networks focus on faces for Rap, Blues, Reggae, R\&B, Latin, and World genres. For Jazz, instruments, typographies and clothes seemed more relevant. A comprehensive feature selection study using evolutionary algorithms explored combinations of audio, semantic tags inferred from audio, MIDI representations, album cover images, playlist co-occurrences, and lyric texts for genre classification \citep{vatolkin2022multi}. For other applications of multi-modal approaches to a whole range of MIR tasks we refer the reader to a recent comprehensive survey \citep{gotham2025towards}.

\section{Discussion}
\label{agenda}

As a result of our analysis of related work in section \ref{stateoftheart} we identify and discuss two major open issues concerning computational research of Pop music culture in section \ref{twoissues} and draft three exemplary promising goals for future work which could overcome these issues in section \ref{threegoals}.

\subsection{Two open issues}
\label{twoissues}

\textbf{Lack of multi-modality:} Our motivational example and the discussion thereof in section \ref{introduction} suggested that Pop music culture needs to be studied as a multi-modal socio-cultural phenomenon. Our survey of MIR articles in section \ref{stateoftheart} however showed that only a small body of work actually does justice to this requirement and provides multi-modal quantitative empirical results about Pop music at scale. Most work we reviewed could in principle be applied to computational analysis of Pop music culture but is so far focused on practical applications. In a more narrow understanding of work that already provides digital and computational musicology at large scale results we like to name \cite{napier2018quantitative,Oramas:etal:2018,brand2019cultural,parada2024song} for distant reading, \cite{Schellenberg:2012,Serra:etal:2012,Mauch:etal:2015,georgieva2024changing,hamilton2024trajectories} for distant hearing, \cite{Schindler:Rauber:2016,Pretet:etal:2021,lopez:flexer:2025} for distant viewing and \citep{nichols2009relationships,McVicarFB11,czedik2024charting,Oramas2025} for multi-modal approaches. Whereas audio and lyrics have been studied to a certain extend, image and video information as well as other text sources on music have received little attention. Therefore we argue that future work on Pop music culture should embrace these under represented as well as different modalities and go beyond studying modalities in isolation.

\noindent
\textbf{Questionable external validity:} In building music corpora, published work reviewed in section \ref{stateoftheart} essentially follows two different routes: either assembling music featured in hit charts or trying to aggregate as much music data as possible. The first approach limits results to music that has been commercially successful in a certain cultural background, which for most studies is the USA. The second approach usually relies on so-called convenience samples \citep{clark2007convenience}, i.e.\ the samples that are drawn from the entirety of all music are dictated by what is available to the researchers rather than on carefully planned and selected samples.
We therefore want to stress the importance of carefully appraising and documenting sampling biases \citep{Fink:2013,Savage:2019} inherent in the sources and building of such music corpora, since such biases might impede external validity of experiments. This is even more important for large music corpora because large sample sizes make statistical significance more likely and hence sampling biases might lead to more spurious significant results \citep{huron2013virtuous}. External validity is the "validity of inferences about the extent to which a causal relationship holds over variations in [...] treatment variables and measurement variables” \citep{shadish:etal:2002}. External validity is the truth of a generalized causal inference drawn from an experiment, i.e.\ which generalizations are justified given certain biases in a music corpus (see \cite{sturm2023review} for an MIR specific discussion). Although it seems clear that sampling biases cannot be completely avoided,  documentation and awareness should help to prevent drawing of unjustified conclusions. Maybe the MIR community can draw on its long and successful history of collaboration with the music industry to gain access to large enough and representative music corpora, as has recently happened with a dataset derived from the industry driven "Music Genome Project" \citep{Oramas2025}.

\subsection{Three goals for future research}
\label{threegoals}

\textbf{Charting the topic universe of Pop music lyrics:} Our first suggested avenue for future research is to extend existing results on hundreds of thousands lyrics \citep{brand2019cultural,parada2024song} to corpora of millions of songs thereby \textbf{strengthening external validity} by providing an overview of the topics present in a truly comprehensive corpus of diverse Pop music lyrics. For topic modeling, a standard unsupervised method is latent Dirichlet allocation (LDA) \citep{latentdiricheltallocation}, but more recent work \citep{bianchi2021pretraining} on Contextualized Topic Modeling allows to also consider the context of words. Adding temporal information (year of publication) will show long range changes in lyrical topics. This could also be complemented by more fine-grained analysis of lyrics, e.g.\ concerning their mood \citep{Delbouys:etal:2018,Hu:Downie:2010,Mishra:etal:2021} or how complex they are \citep{parada2024song}, but also connections to general societal trends could be examined \citep{Oramas:etal:2018}. Connecting topics with meta information like genre will allow to study differences between Pop subcultures, extending previous results on e.g.\ Metal music \cite{czedik2024charting}. An interesting extension to Metal music studies could be exploration of what new topics have developed in lyrics from recently occurring Metal bands with female singers or even all female members\footnote{\url{https://spinditty.com/genres/100-Best-Female-Heavy-Metal-Singers}}. One hypothesis is that the topics are quite different from the misogyny prevalent in newer eras of Metal \cite{kahn2006extreme}.  

\noindent
\textbf{Providing an iconography of Pop album cover art:} Our second proposed avenue of future research is to follow an iconographic approach to study cover art work, \textbf{focusing on a so far under-represented modality}. Iconography in art studies the visual content of artworks to determine their motifs and themes and to characterize the way these are represented, the number of depicted subjects or objects and their mutual relations, trying to interpret their semantic meaning  \citep{milani2021dataset}. Music cover artwork has long been acknowledged to represent a visual genre in its own right and to play an important role in conveying artists' messages and shaping their overall public image \citep{gronstad2010coverscaping}. 

An iconographic approach to a large dataset of album cover art could be achieved trough the use state-of-the-art computer vision technology, building on existing approaches on small data sets \citep{lopez:flexer:2025}, combining automatic image captioning \citep{blip} and zero-shot object detection \citep{groudino}. Whereas image captioning can only recognize objects it has been trained on, which often do not align with what is depicted on album covers, open-set object detection is able to utilize large pre-trained models to detect objects even if they were not part of the labels the model was trained on. Advanced analysis could tackle sexualization of men and women in cover art, extending previous results on Pop magazine covers \citep{hatton2011equal}. Algorithmic analysis of indicators like amount of nudity or body position \citep{lin2021fine} could allow basic quantification of sexualization in millions of album covers instead of thousands of magazine covers as reported before, researching whether the hypothesis of increased sexualization of women, but also men, in recent decades also holds for album cover art. Another interesting aspect is analysis of non-photographic cover art, since object detection algorithms, which usually been trained on photographs, often struggle to recognize objects depicted in different styles such as drawings or paintings, which are also common to a certain extent in album cover art. This is known as the cross-depiction problem and first specialized transfer learning approaches regarding the problem already exist \citep{kadish2021improving}.

\noindent
\textbf{Tracking retro cycles in Pop music's timeline:} Our third suggested topic for future research is to \textbf{use a fully multi-modal approach} to track retro cycles in Pop music. A phenomenon central to Pop music discourse is the fact that trends in music seem to be cyclic, with specific genres becoming popular again after their initial success \citep{Reynolds:2011}. To give one example, mid 1970s Punk rock, with its short, fast songs, simple instrumentation and melodies, was very much influenced by 1960s garage rock \citep{Savage:1991}. After Punk rock diversified into many different niches in the 1980s, that could be subsumed into alternative or indie rock, the 1990s saw a distinct and commercially very successful retro wave of Punk fueled by bands like Green Day or Blink-182. The impact of Punk rock reached all the way into the 21st century with acts like Pussy Riot taking a decidedly political and feminist stance. Similar cycles can be observed for many other genres or popular culture in general \citep{aspers2013sociology}. It has even been hypothesized that this cyclic nature has sped up with the 2000s being "about every other previous decade happening again all at once" \citep{Reynolds:2011}.

These retro cycles are traceable in a range of different modalities. Generally speaking, music genres are characterized by certain highly salient audio properties like instrumentation, rhythm, tempo, use of specific chords, etc, which can be modeled with a range of MIR tools computing respective audio features \citep{Muller:2015}. Another good source of information concerning retro cycles are album reviews, which typically put a new album in the context of previously published music. As one example, the complete digital back catalog of all reviews from the "Rolling Stone" magazine\footnote{\url{https://about.proquest.com/en/products-services/the-rolling-stone-archive/}} is available and could be used for analysis. Even artist's images contain information that could be used to identify retro cycles. Returning to our Punk rock example, its visual culture is characterized by typical haircuts (Mohawk and bright colors) which later have been used by non-Punk artists also. In addition to object recognition approaches presented above, existing specialized approaches for hairstyle detection \citep{muhammad2018hair} could be explored.

All these different modalities should be modeled jointly, building on previous multi-modal approaches \citep{Oramas:etal:2018b,HuangJLGLE22,llark2024} and learning one combined embedding space \citep{Frome:etal:2013}. Distances in the embedding space could be used to compute similarities between songs (or whole albums) which form the basis of distance matrices where the music is ordered according to release date along the axes. Highest similarity will of course be visible along the main axis indicating self-similarity, with larger coherent areas of high similarity constituting a clue for a retro relation. This approach should be formalised by e.g.\ using "Foote novelty" \citep{Foote:2000}, as has been done in related work \citep{Mauch:etal:2015}. Such a multi-modal embedding space could allow joint tracing of influences in different modalities and e.g.\ indicate that artists can be close in their images but not in their audio or textual description. To give just one example, this could show that an artist like Billie Eilish has brightly colored hair, which indicates a Punk influence, but that her music is very different from Punk and that the same holds for her cultural status as evidenced in magazine reviews. This joint multi-modal modeling could therefore allow a more nuanced tracing of musical influences than a purely parallel approach. Additional ablation studies masking individual modalities in the combined embedding space will allow to examine the contribution of different modalities to tracing musical influences.

\section{Conclusion}

Music Information Retrieval has been very successful in powering many commercial but also educational services, reaching from automatic music recommendation to tools helping to learn instruments or assisting in music creation. Although many of these results concern popular music and the analysis of massive amounts of data is at the heart of MIR, our review has shown that a substantial engagement with Pop music as a cultural phenomenon is the exception rather than the rule in MIR's research agenda. We hope that our overview of already existing work, plus the open issues we identified and the exemplary research goals we drafted, will motivate an increase in the empirical and quantitative study of Pop music culture in MIR. This could not only provide a re-orientation for MIR's mainly commercial interest in Pop music, but it would also help to further establish a computational and digital humanities approach to Pop music culture. At the same time researchers need to take existing concerns and criticism of algorithmic approaches to humanities seriously and be conscious about its possible limitations. 

\section*{Funding}

This research was funded in whole by the Austrian Science Fund (FWF) [10.55776/P36653]. For open access purposes, the authors have applied a CC BY public copyright license to any author accepted manuscript version arising from this submission.

\bibliography{biblio}

\end{document}